\documentclass[]{aa}  
\usepackage{graphicx}
\usepackage{txfonts}
\usepackage{hyperref}
\usepackage{orcidlink}
\usepackage{subcaption} 
\usepackage{lscape}  
\usepackage{placeins}
\newcommand {\saxj}{{SAX~J1808.4--3658}\xspace} 
\newcommand {\srga}{{SRGA J144459.2--604207}\xspace} 
\newcommand{\fluxcgs}{erg\,cm$^{-2}$\,s$^{-1}$}

\newcommand {\ixpe}{\textit{IXPE}\xspace}
\newcommand {\nustar}{\textit{NuSTAR}\xspace}

\newcommand {\xmm}{\textit{XMM--Newton}\xspace}

\newcommand {\nicer}{{NICER}\xspace} 
 
\newcommand {\swift}{\textit{Swift}\xspace} 
\newcommand {\extp}{\textit{eXTP}\xspace} 

\newcommand{\simgt}{\stackrel{\displaystyle >}{\,_{\! \,_{\displaystyle
\sim}}}}
\defcitealias{ips2023}{I23}
\defcitealias{pdmp2025}{P25}

\begin{document}

   \title{Polarised X-ray emission from the accreting millisecond pulsar SAX J1808.4--3658 during an X-ray outburst and a type-I X-ray burst}

\authorrunning{Papitto et al.}
\author{
Alessandro Papitto \inst{\ref{in:OAR}}\thanks{        alessandro.papitto@inaf.it}\orcidlink{0000-0001-6289-7413}
\and Alessandro Di Marco \inst{\ref{in:INAF-IAPS}}\orcidlink{0000-0003-0331-3259}
\and Maria Cristina Baglio \inst{\ref{in:OAB}}\orcidlink{0000-0003-1285-4057}
\and Daniele Baldi Tomei \inst{\ref{in:LaSapienza}}
\and Caterina~Ballocco \inst{\ref{in:OAR},\ref{in:LaSapienza}}\orcidlink{0009-0001-0155-7455}
\and Giulia Illiano \inst{\ref{in:CSIC},\ref{in:IEEC},\ref{in:OAB}}\orcidlink{0000-0003-4795-7072}
\and Riccardo~La~Placa \inst{\ref{in:OAR}}\orcidlink{0000-0003-2810-2394}
\and Christian Malacaria \inst{\ref{in:OAR}}\orcidlink{0000-0002-0380-0041}
\and Anna Bobrikova \inst{\ref{in:UTU}}\orcidlink{0009-0009-3183-9742}
\and Juri~Poutanen \inst{\ref{in:UTU}}\orcidlink{0000-0002-0983-0049}
\and Tuomo Salmi \inst{\ref{in:UH}}\orcidlink{0000-0001-6356-125X}
\and Andrea Sanna \inst{\ref{in:UCa}}\orcidlink{0000-0002-0118-2649}
\and Alexandra Veledina \inst{\ref{in:UTU},\ref{in:nordita}}\orcidlink{0000-0002-5767-7253}
\and Filippo Ambrosino \inst{\ref{in:OAR}}\orcidlink{0000-0001-7915-996X}
\and Luciano~Burderi \inst{\ref{in:UCa}}\orcidlink{0000-0001-5458-891X}
\and Sergio Campana \inst{\ref{in:OAB}}\orcidlink{0000-0001-6278-1576}
\and Francesco Coti Zelati \inst{\ref{in:CSIC},\ref{in:IEEC}}\orcidlink{0000-0001-7611-1581}
\and Domitilla de Martino \inst{\ref{in:OANa}}\orcidlink{0000-0002-5069-4202}
\and Tiziana Di Salvo \inst{\ref{in:UPa}}\orcidlink{0000-0002-3220-6375}
\and Gian Luca Israel \inst{\ref{in:OAR}}\orcidlink{0000-0001-5480-6438}
\and Fabio~La~Monaca \inst{\ref{in:UTU}, \ref{in:INAF-IAPS}} \orcidlink{0000-0001-8916-4156}
\and Alessio Marino \inst{\ref{in:USACH},\ref{in:CIRAS},\ref{in:IEEC},\ref{in:CSIC},\ref{in:IASF}}\orcidlink{0000-0001-5674-4664}
\and Arianna Miraval Zanon \inst{\ref{in:ASI}}\orcidlink{0000-0002-0943-4484}
\and Maura~Pilia \inst{\ref{in:OAC}}\orcidlink{0000-0001-7397-8091}
\and Luigi Stella \inst{\ref{in:OAR}}\orcidlink{0000-0002-0018-1687}
}
\institute{
INAF Osservatorio Astronomico di Roma, Via Frascati 33, 00078 Monte Porzio Catone (RM), Italy\label{in:OAR} 
\and 
INAF Istituto di Astrofisica e Planetologia Spaziali, Via del Fosso del Cavaliere 100, 00133 Roma, Italy \label{in:INAF-IAPS}
\and
INAF Osservatorio Astronomico di Brera, Via E. Bianchi 46, 23807 Merate (LC), Italia \label{in:OAB}
\and 
Dipartimento di Fisica, Universit\`{a} degli Studi di Roma ``La Sapienza'', Piazzale Aldo Moro 5, 00185 Roma, Italy \label{in:LaSapienza}
\and
Institute of Space Sciences(ICE, CSIC),Campus UAB, Carrer de Can Magrans s/n, E-08193, Barcelona, Spain \label{in:CSIC}
\and
Institut d’Estudis Espacials de Catalunya (IEEC), E-08860, Castelldefels (Barcelona), Spain \label{in:IEEC}
\and
Department of Physics and Astronomy, 20014 University of Turku, Finland \label{in:UTU}
\and
Department of Physics, P.O. Box 64, 00014 University of Helsinki, Finland \label{in:UH}
\and
Dipartimento di Fisica, Universit\`a degli Studi di Cagliari, SP Monserrato-Sestu, KM 0.7, Monserrato 09042, Italy \label{in:UCa}
\and
Nordita, Stockholm University and KTH Royal Institute of Technology, Hannes Alfvéns väg 12, SE-10691 Stockholm, Sweden \label{in:nordita}
\and
INAF - Osservatorio Astronomico di Capodimonte, Salita Moiariello 16, I-80131 Naples, Italy \label{in:OANa}
\and
Dipartimento di Fisica e Chimica – Emilio Segrè, Università di Palermo, Via Archirafi 36, 90123 Palermo, Italy \label{in:UPa}
\and 
Departamento de Física, Universidad de Santiago de Chile (USACH), Av. Víctor Jara 3493, Estación Central, Chile \label{in:USACH}
\and
Center for Interdisciplinary Research in Astrophysics and Space Sciences (CIRAS), Universidad de Santiago de Chile \label{in:CIRAS}
\and
INAF/IASF Palermo, Via Ugo La Malfa 153, 90146 Palermo, Italy \label{in:IASF}
\and
ASI - Agenzia Spaziale Italiana, Via del Politecnico snc, 00133 Rome (RM), Italy \label{in:ASI}
\and
INAF Osservatorio Astronomico di Cagliari, Via della Scienza 5, 09047 Selargius (CA), Italy \label{in:OAC}
}
\date{Received \today}
\authorrunning{Papitto et al.}
\titlerunning{Polarised X-rays from {\saxj}} 

\abstract{Modelling the X-ray pulse profiles of accreting millisecond pulsars (AMSPs) is a key technique for constraining the masses and radii of neutron stars. To this end, measuring the properties of polarised X-rays provides crucial information on the system geometry. We present the significant detection by \ixpe of linearly polarised 2--8 keV X-rays during the decay of the 2025 outburst of the 2.5~ms AMSP \saxj, with an average polarisation degree of $2.9\pm0.5\%$ and a polarisation angle of $46\degr\pm5\degr$ (measured East of celestial North), offset by $\sim 16^{\degr}$--$24^{\degr}$ from the average direction observed in the optical band over various epochs during the main part of the outburst. The observed degree of X-ray polarisation is similar to that found in the only other source of this class that has been targeted so far. By using a pulse-phase timing solution obtained with \ixpe and \nustar, which confirms the long-term evolution of the pulsar spin and orbit, we did not find any significant pulse phase-resolved variability in the normalised Stokes parameters. Interpreting the polarised X-rays as originating from hot spots on the NS surface indicates they are almost aligned with the spin axis. We also report an indication for linearly polarised emission with a degree of $16\%\pm6\%$ (2--6 keV) during the peak of a type-I thermonuclear X-ray burst, with an angle almost orthogonal to that of the persistent emission. If confirmed, this level would exceed expectations for X-rays reflected by a thin accretion disc. 
}


\maketitle
\nolinenumbers

\section{Introduction}

\begin{figure*}
\centering
\includegraphics[width=0.95\hsize]{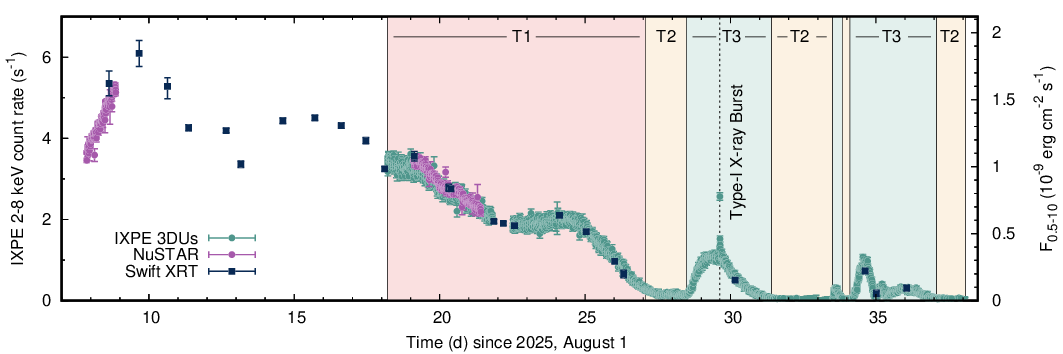}
\caption{Light curve of the 2025 outburst of {\saxj} observed by {\ixpe} (2--8~keV, green dots), {\nustar} (3--79~keV, purple dots) and {\swift}/XRT (0.5--10~keV, dark blue squares). Background subtracted {\ixpe} and {\nustar} data were binned in 300~s intervals and scaled to make the unabsorbed flux seen by each during a 1~day-long interval centred at the epoch of {\swift} observation (2025 August 21, 07:25 UTC, Obs.~Id.~30034178) match the value measured by the {\swift}, $F_{0.5-10}=(8.4\pm0.3)\times10^{-10}$ \fluxcgs. Light red, gold, and green-shaded regions mark the decay (T1), minimum (T2), and reflaring (T3) intervals of the {\ixpe} light curve. A dashed vertical line marks a type-I X-ray burst observed by \ixpe. }
        \label{fig:lcurve}%
    \end{figure*}
    
Modelling the X-ray pulse profiles of rotation-powered millisecond pulsars (MSPs) has recently provided crucial constraints on the masses and radii of neutron stars \citep[NS; see, e.g.][and references therein]{mgs2025,mdh2026}. Accreting millisecond pulsars (AMSPs) offer an intriguing alternative for the scope, as they are orders of magnitude brighter during X-ray outbursts \citep[e.g.,][]{Poutanen2003,snp2018,dsw2025}. Compton scattering by hot (${\sim}50$~keV) electrons in the accretion columns above hot-spots on the NS surface polarises X-ray photons emitted by the surface \citep{vp2004,bls2023}. Modelling the spin phase-resolved polarised radiation provides crucial constraints on geometrical parameters such as the system inclination ($i$) and the co-latitudes of the emitting spots ($\theta_1$ and $\theta_2$), potentially allowing us to obtain similar constraints on the masses and radii of AMSPs to those obtained for rotation-powered MSPs \citep{slk2021,sdw2025}. The {\it Imaging X-ray Polarimetry Explorer} ({\ixpe}) recently detected X-rays with a polarisation degree (PD) of a few per cent from the AMSP {\srga}, suggesting a nearly aligned rotator \citep[][P25 in the following]{pdmp2025} and large spots \citep{dsw2026bis}. Here we present an analysis of an {\ixpe} observation of the prototype of AMSPs, {\saxj} \citep{wvdk1998}, during its 2025 outburst. 

\section{Observations}

On 2025 August 6, the optical monitoring programme of low-mass X-ray binaries 
with the Las Cumbres Observatory \citep[X-ray Binary New Early Warning System;][]{rbl2019} first detected the onset of a new outburst of {\saxj} \citep{ras2025}, later confirmed by follow-up X-ray observations \citep{wmz2025,bpmz2025}. Starting on 2025 August 9, the \swift/X-Ray Telescope (XRT) \citep{Burrows_2005} observed the source (Obs. Id. 19987 and 30034) either in Photon Counting or Windowed Timing modes, depending on the source flux. We created background-subtracted light curves, spectra, and images using the online XRT product builder\footnote{\url{https://www.swift.ac.uk/user_objects/}} \citep{Evans_2009}. Dark blue squares in Fig.~\ref{fig:lcurve} show the unabsorbed 0.5--10~keV flux obtained by using \textsc{xspec} v. 13.0.1 \citep{a96} to model  \swift/XRT spectra with Comptonisation \citep[modelled using the \texttt{thcomp} model, ][]{zsp2020} of blackbody seed photons, absorbed by interstellar matter \citep[modelled with \texttt{tbabs}, ][]{wam2000}. 

 {\ixpe} observed {\saxj} for 918~ks between 2025 August 19 (04:54 UTC) and September 8 (01:41 UTC) within the Target of Opportunity general observer program Id.~04007699 (PI: A.~Papitto). We processed clean data obtained with \texttt{ixpeevtrecon} according to the guide by \citet{dm2026}. We rejected instrumental particle background with \texttt{IXPE-background}\footnote{\url{https://github.com/aledimarco/IXPE-background}} \citep{dmsc2023} employing improved cuts required after the April 2025 detector unit (DU) 2 anomaly. This procedure is now requested regardless of the source brightness to avoid introducing possible systematic effects \citep{dm2026}. We checked the alignment of the three DU images and removed sporadic solar flares characterised by a background rate exceeding 0.4~s$^{-1}$. We extracted source photons from a circular region with a 60\arcsec\ radius around the source position, whereas background photons were extracted from an annular region encompassing the source position with inner and outer radii of 150\arcsec\ and 250\arcsec, respectively. Green dots in Fig.~\ref{fig:lcurve} show the background-subtracted 2--8~keV light curve observed by the three DUs. 
  
We used the contributed script \texttt{ixpe\_protractor}\footnote{\url{https://heasarc.gsfc.nasa.gov/docs/ixpe/analysis/contributed/ixpe_protractor.html}} along with the \ixpe calibration v.~20260424 to get a model-independent measurement of the Stokes parameters $(I,Q,U)$. Except for the interval T2 (see the gold-shaded region in Fig.~\ref{fig:lcurve}), the background was much fainter than the source, and we did not subtract it in the polarimetric analysis \citep{dmsc2023}. We also used \texttt{ixpestartx}\footnote{ \url{https://heasarc.gsfc.nasa.gov/docs/ixpe/analysis/contributed/ixpestartx.html}.} with the \texttt{NEFF} weighted scheme \citep{dmcm22} to produce ancillary and modulation response files, selected the response matrix \texttt{20211209\_alpha075} and binned the resulting spectra to have at least 30 counts per bin. Similar to \citetalias{pdmp2025}, when performing the phase-resolved analysis presented in Sect.~\ref{sec:phaseresolve}, we used the \texttt{pcube} algorithm without acceptance corrections to measure the Stokes parameters (expected to be normally distributed) in counts per second.

The {\it Nuclear Spectroscopic Telescope Array} (\nustar) performed Target of Opportunity observations of \saxj during the outburst rise (Obs. Id. 81101311002) and simultaneously with \ixpe (Id. 81101346002; see purple dots in Fig.~\ref{fig:lcurve}). The latter took place in the context of an \xmm guest observer program (Id. 96132, PI: Papitto). Due to visibility limitations, \xmm (Obs.~Id.~ 0961320101) observation was performed after the end of the \ixpe coverage (Malacaria et al., in prep.). Here, we reduced \nustar data using the calibration version indx20260630, with the fine-clock correction file version 230. We adopted circular regions with a radius of 140$^{\prime\prime}$ around and away from the source position to extract source and background photons in the 3–79 keV band, respectively.

\section{X-ray polarisation analysis}
\label{sec:ixpedata}

\begin{figure}
\centering 
   \includegraphics[width=0.95\hsize]{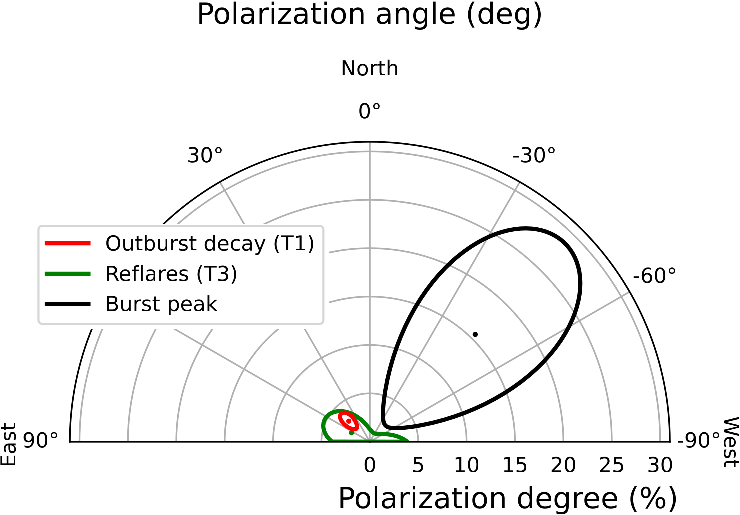}
      \caption{Contours at 90\% c.~l.~ of the 2--6~keV polarised X-ray emission observed by {\ixpe} during the outburst decay (T1, red), reflares (T3, green) and the burst peak (BP, black).  }
         \label{fig:protr}
   \end{figure}

\begin{figure}[t!]
   \centering
   \includegraphics[width=\hsize]{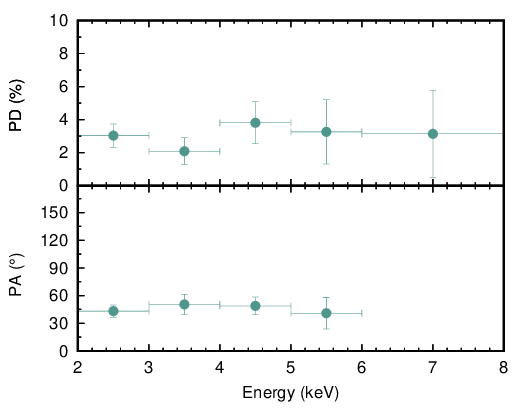}
      \caption{Polarisation degree (top panel) and angle (bottom panel) observed by {\ixpe} during the outburst decay (interval T1; see red-shaded region in Fig.~\ref{fig:lcurve}) as a function of energy. }
         \label{fig:A1}
\end{figure}

{\ixpe} observations (green dots in Fig.~\ref{fig:lcurve}) started halfway through the outburst decay. We split its analysis into decay (T1, red-shaded region in Fig.~\ref{fig:lcurve}), minimum (T2, gold-shaded) and reflaring (T3, green-shaded) intervals. During the interval T1, we significantly measured a model-independent 2--8~keV average polarisation degree (PD), PD$_{\rm T1}=\sqrt{Q^2+U^2}/I=(2.9\pm0.5)$\% (hereafter uncertainties are reported at 68\% c.~l.), at a polarisation angle (PA) of PA$_{\rm T1}=1/2\arctan{(U/Q)}=46\degr\pm5\degr$ (measured east of north, see red countour curves in Fig.~\ref{fig:protr}). We detect polarisation significantly below 6 keV, where we have better statistics. We find no energy dependence otherwise (see Fig.~\ref{fig:A1}). We fitted the 2--8~keV T1 spectrum with the \texttt{thcomp} model to describe Comptonisation by a moderately optically thick ($\tau=1.37\pm0.02$) cloud of $kT_{\rm e}=50$~keV electrons, fully covering $kT_{\rm bb}=0.43\pm0.01$~keV blackbody seed from a region of radius $R_{\rm  bb}=(9.2\pm0.3)\,d_{2.7}$~km (where $d_{2.7}$ is the distance in units of 2.7~kpc, \citealt{ggh2024}), with the hydrogen column density fixed to $N_{\rm H}=1.5\times10^{21}$~cm$^{-2}$ \citep{dssb2019}, giving $\chi^2=490$ for 423 d.~o.~f.  Similar to other instances in this paper, if the model is compatible with the data points, the probability to obtain a value of the $\chi^2$ equal to or higher is $\simgt 1\%$, so this hypothesis cannot be discarded at the 3$\sigma$ confidence level. The spectral model has also been used to perform a spectro-polarimetric (hereafter model-dependent) analysis; fitting the three Stokes parameters with the energy-independent polarisation -- \texttt{polconst} model in \textsc{xspec} -- gave PD$_{\rm T1}=2.8\%\pm0.8\%$ and PA$_{\rm T1}=46\degr\pm8\degr$ with a $\chi^2=794$ for 710~d.o.f., compatible with the model-independent results presented above. During three reflares (interval T3, green-shaded regions in Fig.~\ref{fig:lcurve}) we found similar polarisation properties with a model-independent analysis, PD$_{\rm T3}=2.4\%\pm1.4\%$, PA$_{\rm T3}=65\degr\pm17\degr$ (see green contour curves in Fig.~\ref{fig:protr}), although less significant due to the lower number of counts accumulated. The minimum detectable PD at 99\% confidence level (MDP) of intervals in-between flares (T2, gold-shaded regions in Fig.~\ref{fig:lcurve}) was MDP$_{T2}=14.5$\%, too high to draw firm conclusions.

On 2025 August 30, 15:03 UTC (not corrected to the Solar System barycentre), {\ixpe} also detected a bright \mbox{type-I} X-ray burst during one of the reflares (see the vertical dashed line in Fig.~\ref{fig:lcurve}; \citealt{btp2025}). With a peak count rate of ${\sim}$500~s$^{-1}$, it is the brightest ever seen by {\ixpe}, to our knowledge (see Appendix \ref{app:burst} for details). During a 15~s-long interval around the peak, we observed a 2--6~keV PD$_{\rm BP}=15.5\%\pm6.1\%$, roughly at a right angle compared to the emission seen during the outburst decay (PA$_{\rm BP}=-44\degr\pm11\degr$; see black contour curves in Fig.~\ref{fig:protr}).

\section{Phase-resolved X-ray variability}
\label{sec:phaseresolve}

Starting from the pulsar timing parameters estimated during the last outburst shown by the source in 2022 (see \citep[][I23 hereafter]{ips2023}, we obtained a preliminary timing solution using {\nustar} observations performed during the outburst rise and simultaneously with \ixpe (see purple squares in Fig.~\ref{fig:lcurve} and Appendix \ref{sec:longtermevol} for details). We used this solution to recover the signal in \ixpe $I$ data (see top panel of Fig.~\ref{fig:pulse}); it could be modelled using two harmonic components with fractional amplitudes of $8.9\pm0.2$\% and $0.9\pm0.2$\% (see Eq.~\ref{eq:pulse}). By modelling the evolution of the fundamental harmonic's phase, we obtained a pulse timing solution covering the \ixpe observation, with parameters listed in Table~\ref{table}.  We did not find any significant oscillation in 4~s-long intervals of the type-I burst seen by {\ixpe}, with upper limits on the amplitude at the known frequency $\nu$ during the burst peak in the range 10\%--18\% (3$\sigma$ c.~l.; 16\%--24\% within 5 Hz from $\nu$) much coarser than the values seen from the source in the past \citep[$\simeq 4\%$][]{bjg2019}.

\begin{table}
\caption{\label{table}
Spin and orbital parameters of {\saxj} measured by {\ixpe}.}
\centering
\begin{tabular}{lc}
\hline\hline
\noalign{\smallskip}
\multicolumn{2}{c}{2025 outburst pulse-phase timing} \\
\noalign{\smallskip}
\hline
 $\nu$ (Hz) & $400.97520924 \pm 2\times10^{-8}$ \\
$\ddot{\phi}$ (s$^{-2}$) & $(+2.1\pm0.4)\times10^{-13}$ \\
$a \sin{i}$ (lt-ms) & $62.802\pm0.008$\\
$P_{\rm orb}$ (s) & $7249.154\pm0.005$\\
$T_{\rm asc}$ (MJD) & $60907.052902\pm3\times10^{-6}$\\
Reference epoch (MJD) & 60906.0 \\
\hline
\noalign{\smallskip}
\multicolumn{2}{c}{ Long term evolution (1998--2025)} \\
\noalign{\smallskip}
\hline
 R.A.(J2000) & $18^{\rm h}08^{\rm m}27\fs66\pm0\fs02$\\
 dec (J2000) & $-36\degr58\arcmin44\farcs4\pm0\farcs6$\\
 $\dot{\nu}$ (Hz~s$^{-1}$) & $(-1.21\pm0.09)\times10^{-15}$ \\
$P$ (d) & $8150\pm720$\\
$A$ (s) & $12.4\pm1.3$ \\
$\dot{P}_{\rm orb}$  (s~s$^{-1}$) & $(-4.7\pm1.8)\times10^{-13}$ \\
\hline
\end{tabular}
\end{table}

We used the parameters listed in Table \ref{table} to measure the phase-resolved variability of the 2--6~keV Stokes parameters during the T1 interval (see Fig.~\ref{fig:pulse}). We found no significant variability in the normalised Stokes parameters $Q/I$ and $U/I$, as data points had a 3.9 per cent probability of being drawn from a constant distribution. Yet, we note that dips of the PA appear at pulse phases $0.35$ and $0.5$ (see bottom panel of Fig.~\ref{fig:pulse}), due to a deviation by $\sim 3\sigma$ of the corresponding value of $U/I$ from the average. The low statistics in each of the phase-resolved data points prevented us from assessing the deviation with a higher resolution and understanding if it is structured or is more likely a random fluctuation.

\section{Discussion and conclusions}

The eleventh activity episode since 1996, the 2025 outburst of {\saxj}, confirmed the recurrence time of $\sim$2–3 years (see \citetalias{ips2023} and references therein). The peak flux and light curve morphology closely resemble those of previous events, although {\swift} observations reveal notable post-peak variability, which is further supported by higher-cadence {\it Einstein Probe} observations (Illiano et al., in prep.). The measured spin and orbital parameters follow the established long-term trends (\citetalias{ips2023}). Pending confirmation by future observations, gravitational coupling between the orbital motion and variations in the donor's quadrupole moment offers a viable explanation for the $\sim$22~yr-long oscillation in the epoch of passage of the pulsar at the ascending node (see Fig.~\ref{fig:spinorbevol}; \citealt{pjk2017,sdsb2017}).

\begin{figure}
\centering
\includegraphics[width=0.9\hsize]{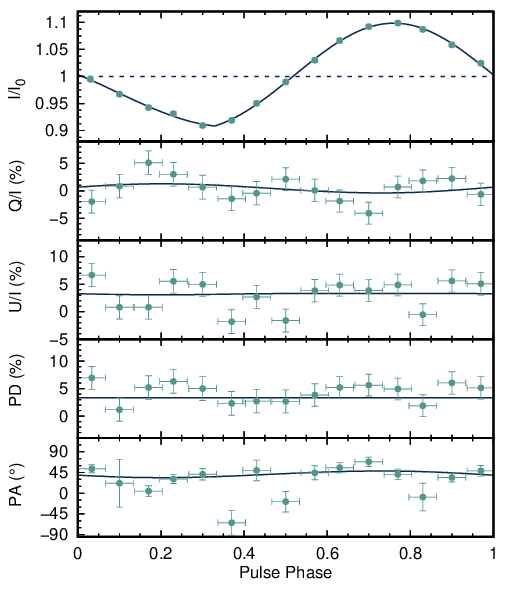}
\caption{Pulse-phase-resolved variability of Stokes parameters. 
From top to bottom: Count rate $I$ normalised to the average $I_0=2.38$~s$^{-1}$, normalised Stokes parameters $Q/I$ and $U/I$, and the PD and PA observed by \ixpe in the 2--6~keV band during the T1 interval using 15 spin-phase bins. Solid lines indicate the model with two point-like spots that best fits the observed Stokes parameter pulse profiles (see Appendix~ \ref{sec:pulsemodel}).}
         \label{fig:pulse}
   \end{figure}
   
During the outburst decay, {\ixpe} detected linearly polarised 2--8~keV X-rays at an average degree of $2.9\%\pm0.5\%$, similar to {\srga} (\citetalias{pdmp2025}). Unlike the marked decrease of the PD below 3~keV observed in the latter, {\saxj} showed no appreciable energy dependence. The X-ray PD of the two AMSPs targeted by {\ixpe} thus far is consistent with Comptonization by mildly relativistic electrons in the accretion shock \citep{bls2023}. Recent relativistic ray-tracing analyses of the X-ray pulse profiles of {\saxj} \citep{dsw2026} and {\srga} \citep{dsw2026bis}, incorporating accretion-disc emission \citep{dsw2025} and, in the latter case, disc obscuration \citep{mdw2026}, suggest the presence of very large hot spots with an aperture ranging from a few tens of degrees up to half the surface area. The expected polarisation for such a configuration is significantly lower than for point-like spots \citep{sdw2025}, helping to account for the mild linear X-ray polarisation observed in these two AMSPs.

As with \srga (\citetalias{pdmp2025}, \citealt{dsw2026bis}), emitting spots closely aligned with the spin axis may account for the lack of significant phase-resolved variability in the Stokes parameters. Modelling the pulse profiles of Stokes parameters $(I,Q,U)$ of \saxj using the simplistic approach of \citetalias{pdmp2025}, which assumes two point-like spots on the surface of a spherical NS, constant radiation anisotropy and energy-independent polarisation, yields spot co-latitudes equal to $\theta_1=6{\degr}^{\,{+2{\degr}}}_{\,-1{\degr}}$ and $\theta_2=177{\degr}\pm 1{\degr}$ (see Fig.~\ref{fig:pulse} and Appendix~\ref{sec:pulsemodel} for details). Employing a full spectropolarimetric framework that incorporates wide spots, a realistic treatment of  Compton scattering, and accretion disc emission and obscuration \citep{slk2021,sdw2025,bls2023,dsw2025,dsw2026,mdw2026} is essential to confirm these preliminary indications. We note that both approaches gave similar estimates of $\theta_1$ and $\theta_2$ for \srga \citep{dsw2026bis}. Such a configuration implies a close alignment between the magnetic field and the spin axis, a scenario previously invoked to explain the generally small and variable oscillation amplitudes observed from AMSPs, the variability of their pulse shapes and phases, as well as the pulse intermittence observed from a few of them \citep{lbvw2009,lbvw2009bis}. 

For a moderately optically thick medium, as inferred for \saxj ($\tau\simeq 1.4$), the X-ray polarisation produced by Compton scattering in a plane-parallel slab over the surface is expected to align with the slab axis \citep{st1985,ps1996,bls2023}. Consequently, the polarisation vector revolves around the direction of the NS spin axis, which in turn is aligned with the disc/binary orbital axis as a consequence of the prolonged accretion history that brought an AMSP such as \saxj to its current spin period. The average X-ray PA during the outburst decay (PA$_{\rm T1}=1/2\arctan{(U/Q)}=46\degr\pm5\degr$) lies not far from the optical linear polarisation, whose PA varies on sub-hour timescales but clusters around a mean of $\sim30\degr$ both in 2019 \citep{brc2020} and in 2025 outbursts, including VLT observations simultaneous with the \ixpe coverage \citep{bip26}. Unlike the X-ray emission, polarised optical photons observed during the outburst decay likely originate from a jet, as the rapid variability of the optical PA disfavours scattering in the disc. The average optical PA measured in the $\it R$-band and {\it I}-band during the main part of the outburst is offset by $11^{\degr} \pm 5^{\degr}$  and $24^{\degr} \pm 5^{\degr}$  from the X-ray one, respectively. \citet{bip26} ascribed such a polarised optical emission to originate from a wide jet base, while the collimated jet remains aligned with the spin/binary axis. At the onset of the reflaring stage, instead, the optical PA coincides with the X-ray one, as expected if the optical polarised emission flux is dominated by emission from the disc–magnetosphere interaction region, also proposed to produce the optical pulsations of this source \citep{bmp2026}, or by scattering in an outflow.

The indication of linearly polarised X-ray emission significant at the 2.5$\sigma$ level, ${\rm PD}_{\rm BP}=(16\pm6)\%$, at the peak of a type-I X-ray burst is striking. For comparison, \citetalias{pdmp2025} obtained an upper limit of 8.5\% (90\% c.l.) on the PD from 52 stacked bursts from \srga. If confirmed by future observations, the indication of enhanced polarisation during the burst could be attributed to the scattering or reflection of the incident emission, driven by a change in either the angular profile of the incident emission or the geometry of the scatterer/reflector relative to the persistent state. For nearly isotropic burst emission, the reflection of NS radiation off the accretion disc can produce ${\rm PD}\sim3$--4\% for $i=60\degr$--$80\degr$ \citep{ls1985,tdo2026}. Alternatively, scattering off disc wind can produce enhanced polarisation \citep{nvp2025}. However, in both cases the polarisation angle is expected to be aligned with the system's symmetry axis.

The observed polarisation angle provides an additional clue regarding the origin of the burst polarisation suggested by \ixpe data. The PA at the burst peak is roughly orthogonal to that of the persistent emission (Fig.~\ref{fig:protr}), suggesting a burst polarisation orientation orthogonal to the disc axis. Such an orientation can arise from scattering in a vertically extended structure or an extended scattering region inflated by the pressure of the burst emission. For instance, scattering in an accretion flow can produce a PD of up to $\sim60\%$ at $i=60\degr$ in the single-scattering regime. If the scattered component contributes roughly $20\%$ of the total flux, a level expected for a low optical depth that avoids depolarisation from multiple scatterings, the resulting net PD is $ \sim 12\%$, consistent with the observed value. An even higher PD could naturally arise if the incident source is obscured, as in the case of Cyg X-3 \citep{vmp2024,vpb2024}. Interestingly, reflection features at $1$ and $6.7$~keV alongside ratios of the reflected flux by the disc over the directly observed burst blackbody emission exceeding unity have been reported during type-I X-ray bursts from {\saxj} \citep{izgm2013,bjg2019,mnc2026}, suggesting partial obscuration of the source also in this case \citep{hk2016}.

\ixpe observations of two AMSPs have revealed linear X-ray polarisation at the $2\%$--$3\%$ level, with no detectable phase-resolved variability. This suggests a configuration with the emitting spots closely aligned with the spin axis. Future \ixpe campaigns will determine whether aligned rotators are a common feature among AMSPs. While such a geometry limits the precision with which pulsar parameters can be inferred from polarisation data \citep{sdw2025}, it underscores the transformative potential of large effective area X-ray detectors of upcoming missions like {\extp} for using AMSPs to constrain the NS equation of state \citep{zsx2025,lwz2025}.
Future X-ray polarimetric observations will also clarify the feasibility of probing the impact of type-I X-ray bursts on accretion inflows.

\begin{acknowledgements}
We thank the {\ixpe}, {\nustar} and {\swift} teams for scheduling these ToO observations.
This work was supported by INAF Research Grants ``PULSE-X'' and ``Chasing the optical counterpart in FRBs and Magnetars with SiFAP2@TNG and SiFAP-SoFT@Cassini'', and by Fond.~Cariplo/CDP through Grant 2023-2560. AB, JP, AV, and FLM were supported by the Research Council of Finland grants 355672 and 372881 and Centre of Excellence in Neutron-Star Physics (grant 374064). Nordita is supported in part by NordForsk. TS acknowledges funding by the Research Council of Finland grant 368807 and the Centre of Excellence in Neutron-Star Physics (grant 374063). AM acknowledges support from the Fund Vera Rubin/Chile 2024, under the project DIA 1736 ``Silent black holes around red supergiants''. GI is supported by a Juan de la Cierva fellowship (JDC2024-053550-I). This work made use of data supplied by the UK Swift Science Data Centre at the University of Leicester.
\end{acknowledgements}

\bibliographystyle{aa} 
\bibliography{ref.bib}

\begin{thebibliography}{56}
\expandafter\ifx\csname natexlab\endcsname\relax\def\natexlab#1{#1}\fi

\bibitem[{{Arnaud}(1996)}]{a96}
{Arnaud}, K.~A. 1996, in Astronomical Society of the Pacific Conference Series, Vol. 101, Astronomical Data Analysis Software and Systems V, ed. G.~H. {Jacoby} \& J.~{Barnes}, 17

\bibitem[{Baglio {et~al.}(2026)Baglio, Illiano, Papitto, Russell, Marco, Alabarta, Ballocco, Campana, Covino, D'Avanzo, Fender, Gasealahwe, Lewis, Messa, Motta, \& Saikia}]{bip26}
Baglio, M.~C., Illiano, G., Papitto, A., {et~al.} 2026, \aap, in press [\eprint[arXiv]{2611.00221}]

\bibitem[{{Baglio} {et~al.}(2020){Baglio}, {Russell}, {Crespi}, {Covino}, {Johar}, {Homan}, {Bramich}, {Saikia}, {Campana}, {D'Avanzo}, {Fender}, {Goldoni}, {Goodwin}, {Lewis}, {Masetti}, {Miraval Zanon}, {Motta}, {Mu{\~n}oz-Darias}, \& {Shahbaz}}]{brc2020}
{Baglio}, M.~C., {Russell}, D.~M., {Crespi}, S., {et~al.} 2020, \apj, 905, 87

\bibitem[{{Ballocco} {et~al.}(2026){Ballocco}, {Miraval Zanon}, {Papitto}, {Burgay}, {Baglio}, {Malacaria}, {Ambrosino}, {Possenti}, {Illiano}, {Barr}, {Breton}, {Burderi}, {Campana}, {Clark}, {Coti Zelati}, {de Martino}, {Di Salvo}, {Kramer}, {La Placa}, {Lewis}, {Marino}, {Russell}, {Sanna}, {Stella}, {Stappers}, {Turner}, {Cecconi}, \& {Ghedina}}]{bmp2026}
{Ballocco}, C., {Miraval Zanon}, A., {Papitto}, A., {et~al.} 2026, \aap, in press [\eprint[arXiv]{2610.18322}]

\bibitem[{{Ballocco} {et~al.}(2025{\natexlab{a}}){Ballocco}, {Papitto}, {Zanon}, {Illiano}, \& {Ambrosino}}]{bpmz2025}
{Ballocco}, C., {Papitto}, A., {Zanon}, A.~M., {Illiano}, G., \& {Ambrosino}, F. 2025{\natexlab{a}}, ATel, 17326, 1

\bibitem[{{Ballocco} {et~al.}(2025{\natexlab{b}}){Ballocco}, {Trois}, {Papitto}, {Del Monte}, {Di Marco}, {La Monaca}, {Illiano}, {Ambrosino}, \& {Miraval Zanon}}]{btp2025}
{Ballocco}, C., {Trois}, A., {Papitto}, A., {et~al.} 2025{\natexlab{b}}, ATel, 17369, 1

\bibitem[{{Bobrikova} {et~al.}(2023){Bobrikova}, {Loktev}, {Salmi}, \& {Poutanen}}]{bls2023}
{Bobrikova}, A., {Loktev}, V., {Salmi}, T., \& {Poutanen}, J. 2023, \aap, 678, A99

\bibitem[{{Bult} {et~al.}(2020){Bult}, {Chakrabarty}, {Arzoumanian}, {Gendreau}, {Guillot}, {Malacaria}, {Ray}, \& {Strohmayer}}]{bca2020}
{Bult}, P., {Chakrabarty}, D., {Arzoumanian}, Z., {et~al.} 2020, \apj, 898, 38

\bibitem[{{Bult} {et~al.}(2019){Bult}, {Jaisawal}, {G{\"u}ver}, {Strohmayer}, {Altamirano}, {Arzoumanian}, {Ballantyne}, {Chakrabarty}, {Chenevez}, {Gendreau}, {Guillot}, \& {Ludlam}}]{bjg2019}
{Bult}, P., {Jaisawal}, G.~K., {G{\"u}ver}, T., {et~al.} 2019, \apjl, 885, L1

\bibitem[{{Burrows} {et~al.}(2005){Burrows}, {Hill}, {Nousek}, {Kennea}, {Wells}, {Osborne}, {Abbey}, {Beardmore}, {Mukerjee}, {Short}, {Chincarini}, {Campana}, {Citterio}, {Moretti}, {Pagani}, {Tagliaferri}, {Giommi}, {Capalbi}, {Tamburelli}, {Angelini}, {Cusumano}, {Br{\"a}uninger}, {Burkert}, \& {Hartner}}]{Burrows_2005}
{Burrows}, D.~N., {Hill}, J.~E., {Nousek}, J.~A., {et~al.} 2005, \ssr, 120, 165

\bibitem[{{Di Marco}(2026, to appear in X-ray Polarimetry: Detection, Observations, Modeling and the Future, ed. H.~Liu, A.~Ingram (Springer, Singapore))}]{dm2026}
{Di Marco}, A. 2026, to appear in X-ray Polarimetry: Detection, Observations, Modeling and the Future, ed. H.~Liu, A.~Ingram (Springer, Singapore) [\eprint[arXiv]{2604.03366}]

\bibitem[{{Di Marco} {et~al.}(2022){Di Marco}, {Costa}, {Muleri}, {Soffitta}, {Fabiani}, {La Monaca}, {Rankin}, {Xie}, {Bachetti}, {Baldini}, {Baumgartner}, {Bellazzini}, {Brez}, {Castellano}, {Del Monte}, {Di Lalla}, {Ferrazzoli}, {Latronico}, {Maldera}, {Manfreda}, {O'Dell}, {Perri}, {Pesce-Rollins}, {Puccetti}, {Ramsey}, {Ratheesh}, {Sgr{\`o}}, {Spandre}, {Tennant}, {Tobia}, {Trois}, \& {Weisskopf}}]{dmcm22}
{Di Marco}, A., {Costa}, E., {Muleri}, F., {et~al.} 2022, \aj, 163, 170

\bibitem[{{Di Marco} {et~al.}(2023){Di Marco}, {Soffitta}, {Costa}, {Ferrazzoli}, {La Monaca}, {Rankin}, {Ratheesh}, {Xie}, {Baldini}, {Del Monte}, {Ehlert}, {Fabiani}, {Kim}, {Muleri}, {O'Dell}, {Ramsey}, {Rubini}, {Sgr{\`o}}, {Silvestri}, {Tennant}, \& {Weisskopf}}]{dmsc2023}
{Di Marco}, A., {Soffitta}, P., {Costa}, E., {et~al.} 2023, \aj, 165, 143

\bibitem[{{Di Salvo} {et~al.}(2019){Di Salvo}, {Sanna}, {Burderi}, {Papitto}, {Iaria}, {Gambino}, \& {Riggio}}]{dssb2019}
{Di Salvo}, T., {Sanna}, A., {Burderi}, L., {et~al.} 2019, \mnras, 483, 767

\bibitem[{{Dorsman} {et~al.}(2026{\natexlab{a}}){Dorsman}, {Salmi}, {Watts}, {Ng}, {Bobrikova}, {Loktev}, {Poutanen}, \& {Wilms}}]{dsw2026}
{Dorsman}, B., {Salmi}, T., {Watts}, A.~L., {et~al.} 2026{\natexlab{a}}, \mnras, 545, staf1983

\bibitem[{{Dorsman} {et~al.}(2026{\natexlab{b}}){Dorsman}, {Salmi}, {Watts}, {Ng}, {Bobrikova}, {Marco}, {Galloway}, {Guillot}, {Hoogkamer}, {Kini}, {Monaca}, {Loktev}, {Lucchini}, {Malacaria}, {Mao}, {Papitto}, \& {Poutanen}}]{dsw2026bis}
{Dorsman}, B., {Salmi}, T., {Watts}, A.~L., {et~al.} 2026{\natexlab{b}}, \mnras, 551, stag1340

\bibitem[{{Dorsman} {et~al.}(2025){Dorsman}, {Salmi}, {Watts}, {Ng}, {Kamath}, {Bobrikova}, {Poutanen}, {Loktev}, {Kini}, {Choudhury}, {Vinciguerra}, {Bogdanov}, \& {Chakrabarty}}]{dsw2025}
{Dorsman}, B., {Salmi}, T., {Watts}, A.~L., {et~al.} 2025, \mnras, 538, 2853

\bibitem[{{Evans} {et~al.}(2009){Evans}, {Beardmore}, {Page}, {Osborne}, {O'Brien}, {Willingale}, {Starling}, {Burrows}, {Godet}, {Vetere}, {Racusin}, {Goad}, {Wiersema}, {Angelini}, {Capalbi}, {Chincarini}, {Gehrels}, {Kennea}, {Margutti}, {Morris}, {Mountford}, {Pagani}, {Perri}, {Romano}, \& {Tanvir}}]{Evans_2009}
{Evans}, P.~A., {Beardmore}, A.~P., {Page}, K.~L., {et~al.} 2009, \mnras, 397, 1177

\bibitem[{{Galloway} {et~al.}(2024){Galloway}, {Goodwin}, {Hilder}, {Waterson}, \& {Cup{\'a}k}}]{ggh2024}
{Galloway}, D.~K., {Goodwin}, A.~J., {Hilder}, T., {Waterson}, L., \& {Cup{\'a}k}, M. 2024, \mnras, 535, 647

\bibitem[{{Hartman} {et~al.}(2008){Hartman}, {Patruno}, {Chakrabarty}, {Kaplan}, {Markwardt}, {Morgan}, {Ray}, {van der Klis}, \& {Wijnands}}]{hpc2008}
{Hartman}, J.~M., {Patruno}, A., {Chakrabarty}, D., {et~al.} 2008, \apj, 675, 1468

\bibitem[{{He} \& {Keek}(2016)}]{hk2016}
{He}, C.-C. \& {Keek}, L. 2016, \apj, 819, 47

\bibitem[{{Illiano} {et~al.}(2023){Illiano}, {Papitto}, {Sanna}, {Bult}, {Ambrosino}, {Miraval Zanon}, {Coti Zelati}, {Stella}, {Altamirano}, {Baglio}, {Bozzo}, {Burderi}, {de Martino}, {Di Marco}, {di Salvo}, {Ferrigno}, {Loktev}, {Marino}, {Ng}, {Pilia}, {Poutanen}, \& {Salmi}}]{ips2023}
{Illiano}, G., {Papitto}, A., {Sanna}, A., {et~al.} 2023, \apjl, 942, L40

\bibitem[{{in't Zand} {et~al.}(2013){in't Zand}, {Galloway}, {Marshall}, {Ballantyne}, {Jonker}, {Paerels}, {Palmer}, {Patruno}, \& {Weinberg}}]{izgm2013}
{in't Zand}, J.~J.~M., {Galloway}, D.~K., {Marshall}, H.~L., {et~al.} 2013, \aap, 553, A83

\bibitem[{{Lamb} {et~al.}(2009{\natexlab{a}}){Lamb}, {Boutloukos}, {Van Wassenhove}, {Chamberlain}, {Lo}, {Clare}, {Yu}, \& {Miller}}]{lbvw2009bis}
{Lamb}, F.~K., {Boutloukos}, S., {Van Wassenhove}, S., {et~al.} 2009{\natexlab{a}}, \apj, 706, 417

\bibitem[{{Lamb} {et~al.}(2009{\natexlab{b}}){Lamb}, {Boutloukos}, {Van Wassenhove}, {Chamberlain}, {Lo}, \& {Miller}}]{lbvw2009}
{Lamb}, F.~K., {Boutloukos}, S., {Van Wassenhove}, S., {et~al.} 2009{\natexlab{b}}, \apjl, 705, L36

\bibitem[{{Lapidus} \& {Sunyaev}(1985)}]{ls1985}
{Lapidus}, I.~I. \& {Sunyaev}, R.~A. 1985, \mnras, 217, 291

\bibitem[{{Li} {et~al.}(2025){Li}, {Watts}, {Zhang}, {Guillot}, {Xu}, {Santangelo}, {Zane}, {Feng}, {Zhang}, {Ge}, {Qi}, {Salmi}, {Dorsman}, {Miao}, {Tu}, {Cavecchi}, {Zhou}, {Zheng}, {Wang}, {Cheng}, {Liu}, {Wei}, {Wang}, {Xu}, {Weng}, {Zhu}, {Li}, {Shao}, {Tuo}, {Dohi}, {Lyu}, {Liu}, {Yuan}, {Wang}, {Zhang}, {Li}, {Tao}, {Zhang}, {Shen}, {Provid{\^e}ncia}, {Tolos}, {Patruno}, {Li}, {Liu}, {Zhou}, {Chen}, {Fan}, {Kajino}, {Lai}, {Li}, {Meng}, {Tang}, {Xiao}, {Xiong}, {Xu}, {Zhou}, {Ballantyne}, {Burgio}, {Chenevez}, {Choudhury}, {Fantina}, {Galloway}, {Gulminelli}, {Hebeler}, {Hoogkamer}, {Horvath}, {Kini}, {Kurkela}, {Linares}, {Margueron}, {Mendes}, {Oertel}, {Papitto}, {Poutanen}, {Rea}, {Schwenk}, {Song}, {Svensson}, {Tsang}, {Vuorinen}, {Andersson}, {Miller}, {Rezzolla}, {Stone}, \& {Thomas}}]{lwz2025}
{Li}, A., {Watts}, A.~L., {Zhang}, G., {et~al.} 2025, Science China Physics, Mechanics, and Astronomy, 68, 119503

\bibitem[{{Mandal} {et~al.}(2026){Mandal}, {Naik}, \& {Chhotaray}}]{mnc2026}
{Mandal}, M., {Naik}, S., \& {Chhotaray}, B. 2026, \mnras, 548, stag624

\bibitem[{{Mao} {et~al.}(2026){Mao}, {Dorsman}, {Watts}, {Salmi}, {Poutanen}, \& {Li}}]{mdw2026}
{Mao}, Y.-H., {Dorsman}, B., {Watts}, A.~L., {et~al.} 2026, \mnras, in press [\eprint[arXiv]{2606.14493}]

\bibitem[{{Mauviard} {et~al.}(2025){Mauviard}, {Guillot}, {Salmi}, {Choudhury}, {Dorsman}, {Gonz{\'a}lez-Caniulef}, {Hoogkamer}, {Huppenkothen}, {Kazantsev}, {Kini}, {Olive}, {Stammler}, {Watts}, {Mendes}, {Rutherford}, {Schwenk}, {Svensson}, {Bogdanov}, {Kerr}, {Ray}, {Guillemot}, {Cognard}, \& {Theureau}}]{mgs2025}
{Mauviard}, L., {Guillot}, S., {Salmi}, T., {et~al.} 2025, \apj, 995, 60

\bibitem[{{Miller} {et~al.}(2026){Miller}, {Dittmann}, {Holt}, {Lamb}, {Chirenti}, {Arzoumanian}, {Berteaud}, {Bogdanov}, {Gendreau}, {Ho}, {Morsink}, {Ray}, {Remillard}, {Wadiasingh}, \& {Wolff}}]{mdh2026}
{Miller}, M.~C., {Dittmann}, A.~J., {Holt}, I.~M., {et~al.} 2026, \apjl, 1000, L48

\bibitem[{Newville {et~al.}(2025)Newville, Otten, Nelson, Stensitzki, Ingargiola, Allan, Fox, Carter, \& Rawlik}]{lmfit}
Newville, M., Otten, R., Nelson, A., {et~al.} 2025, LMFIT: Non-Linear Least-Squares Minimization and Curve-Fitting for Python

\bibitem[{{Nitindala} {et~al.}(2025){Nitindala}, {Veledina}, \& {Poutanen}}]{nvp2025}
{Nitindala}, A.~P., {Veledina}, A., \& {Poutanen}, J. 2025, \aap, 694, A230

\bibitem[{{Papitto} {et~al.}(2025){Papitto}, {Di Marco}, {Poutanen}, {Salmi}, {Illiano}, {La Monaca}, {Ambrosino}, {Bobrikova}, {Baglio}, {Ballocco}, {Burderi}, {Campana}, {Coti Zelati}, {Di Salvo}, {La Placa}, {Loktev}, {Long}, {Malacaria}, {Miraval Zanon}, {Ng}, {Pilia}, {Sanna}, {Stella}, {Strohmayer}, \& {Zane}}]{pdmp2025}
{Papitto}, A., {Di Marco}, A., {Poutanen}, J., {et~al.} 2025, \aap, 694, A37

\bibitem[{{Patruno} {et~al.}(2017){Patruno}, {Jaodand}, {Kuiper}, {Bult}, {Hessels}, {Knigge}, {King}, {Wijnands}, \& {van der Klis}}]{pjk2017}
{Patruno}, A., {Jaodand}, A., {Kuiper}, L., {et~al.} 2017, \apj, 841, 98

\bibitem[{{Poutanen}(2020)}]{p2020}
{Poutanen}, J. 2020, \aap, 641, A166

\bibitem[{{Poutanen} \& {Beloborodov}(2006)}]{pb2006}
{Poutanen}, J. \& {Beloborodov}, A.~M. 2006, \mnras, 373, 836

\bibitem[{{Poutanen} \& {Gierli{\'n}ski}(2003)}]{Poutanen2003}
{Poutanen}, J. \& {Gierli{\'n}ski}, M. 2003, \mnras, 343, 1301

\bibitem[{{Poutanen} \& {Svensson}(1996)}]{ps1996}
{Poutanen}, J. \& {Svensson}, R. 1996, \apj, 470, 249

\bibitem[{{Russell} {et~al.}(2025){Russell}, {Alabarta}, {Saikia}, {Rout}, {Bramich}, {Baglio}, {Illiano}, {Russell}, \& {Lewis}}]{ras2025}
{Russell}, D.~M., {Alabarta}, K., {Saikia}, P., {et~al.} 2025, ATel, 17323, 1

\bibitem[{{Russell} {et~al.}(2019){Russell}, {Bramich}, {Lewis}, {AlMannaei}, {Al Qaissieh}, {Al Qasim}, {Al Yazeedi}, {Baglio}, {Bernardini}, {Elgalad}, {Gabuya}, {Lasota}, {Palado}, {Roche}, {Shivkumar}, {Udrescu}, \& {Zhang}}]{rbl2019}
{Russell}, D.~M., {Bramich}, D.~M., {Lewis}, F., {et~al.} 2019, Astronomische Nachrichten, 340, 278

\bibitem[{{Salmi} {et~al.}(2025){Salmi}, {Dorsman}, {Watts}, {Bobrikova}, {Marco}, {Loktev}, {Papitto}, {Pilia}, {Poutanen}, \& {Rankin}}]{sdw2025}
{Salmi}, T., {Dorsman}, B., {Watts}, A.~L., {et~al.} 2025, \mnras, 538, 2562

\bibitem[{{Salmi} {et~al.}(2021){Salmi}, {Loktev}, {Korsman}, {Baldini}, {Tsygankov}, \& {Poutanen}}]{slk2021}
{Salmi}, T., {Loktev}, V., {Korsman}, K., {et~al.} 2021, \aap, 646, A23

\bibitem[{{Salmi} {et~al.}(2018){Salmi}, {N{\"a}ttil{\"a}}, \& {Poutanen}}]{snp2018}
{Salmi}, T., {N{\"a}ttil{\"a}}, J., \& {Poutanen}, J. 2018, \aap, 618, A161

\bibitem[{{Sanna} {et~al.}(2017){Sanna}, {Di Salvo}, {Burderi}, {Riggio}, {Pintore}, {Gambino}, {Iaria}, {Tailo}, {Scarano}, \& {Papitto}}]{sdsb2017}
{Sanna}, A., {Di Salvo}, T., {Burderi}, L., {et~al.} 2017, \mnras, 471, 463

\bibitem[{{Suleimanov} {et~al.}(2023){Suleimanov}, {Forsblom}, {Tsygankov}, {Poutanen}, {Doroshenko}, {Doroshenko}, {Capitanio}, {Di Marco}, {Gonz{\'a}lez-Caniulef}, {Heyl}, {La Monaca}, {Lutovinov}, {Molkov}, {Malacaria}, {Mushtukov}, {Shtykovsky}, {Agudo}, {Antonelli}, {Bachetti}, {Baldini}, {Baumgartner}, {Bellazzini}, {Bianchi}, {Bongiorno}, {Bonino}, {Brez}, {Bucciantini}, {Castellano}, {Cavazzuti}, {Chen}, {Ciprini}, {Costa}, {De Rosa}, {Del Monte}, {Di Gesu}, {Di Lalla}, {Donnarumma}, {Dov{\v{c}}iak}, {Ehlert}, {Enoto}, {Evangelista}, {Fabiani}, {Ferrazzoli}, {Garcia}, {Gunji}, {Hayashida}, {Iwakiri}, {Jorstad}, {Kaaret}, {Karas}, {Kislat}, {Kitaguchi}, {Kolodziejczak}, {Krawczynski}, {Latronico}, {Liodakis}, {Maldera}, {Manfreda}, {Marin}, {Marinucci}, {Marscher}, {Marshall}, {Massaro}, {Matt}, {Mitsuishi}, {Mizuno}, {Muleri}, {Negro}, {Ng}, {O'Dell}, {Omodei}, {Oppedisano}, {Papitto}, {Pavlov}, {Peirson}, {Perri}, {Pesce-Rollins}, {Petrucci}, {Pilia}, {Possenti}, {Puccetti}, {Ramsey}, {Rankin},
  {Ratheesh}, {Roberts}, {Romani}, {Sgr{\`o}}, {Slane}, {Soffitta}, {Spandre}, {Swartz}, {Tamagawa}, {Tavecchio}, {Taverna}, {Tawara}, {Tennant}, {Thomas}, {Tombesi}, {Trois}, {Turolla}, {Vink}, {Weisskopf}, {Wu}, {Xie}, \& {Zane}}]{sft2023}
{Suleimanov}, V.~F., {Forsblom}, S.~V., {Tsygankov}, S.~S., {et~al.} 2023, \aap, 678, A119

\bibitem[{{Sunyaev} \& {Titarchuk}(1985)}]{st1985}
{Sunyaev}, R.~A. \& {Titarchuk}, L.~G. 1985, \aap, 143, 374

\bibitem[{{Tomaru} {et~al.}(2026){Tomaru}, {Done}, \& {Odaka}}]{tdo2026}
{Tomaru}, R., {Done}, C., \& {Odaka}, H. 2026, \mnras, 547, stag498

\bibitem[{{Veledina} {et~al.}(2024{\natexlab{a}}){Veledina}, {Muleri}, {Poutanen}, {Podgorn{\'y}}, {Dov{\v{c}}iak}, {Capitanio}, {Churazov}, {De Rosa}, {Di Marco}, {Forsblom}, {Kaaret}, {Krawczynski}, {La Monaca}, {Loktev}, {Lutovinov}, {Molkov}, {Mushtukov}, {Ratheesh}, {Rodriguez Cavero}, {Steiner}, {Sunyaev}, {Tsygankov}, {Weisskopf}, {Zdziarski}, {Bianchi}, {Bright}, {Bursov}, {Costa}, {Egron}, {Garcia}, {Green}, {Gurwell}, {Ingram}, {Kajava}, {Kale}, {Kraus}, {Malyshev}, {Marin}, {Matt}, {McCollough}, {Mereminskiy}, {Nizhelsky}, {Piano}, {Pilia}, {Pittori}, {Rao}, {Righini}, {Soffitta}, {Shevchenko}, {Svoboda}, {Tombesi}, {Trushkin}, {Tsybulev}, {Ursini}, {Wu}, {Agudo}, {Antonelli}, {Bachetti}, {Baldini}, {Baumgartner}, {Bellazzini}, {Bongiorno}, {Bonino}, {Brez}, {Bucciantini}, {Castellano}, {Cavazzuti}, {Chen}, {Ciprini}, {Del Monte}, {Di Gesu}, {Di Lalla}, {Donnarumma}, {Doroshenko}, {Ehlert}, {Enoto}, {Evangelista}, {Fabiani}, {Ferrazzoli}, {Gunji}, {Hayashida}, {Heyl}, {Iwakiri}, {Jorstad}, {Karas},
  {Kislat}, {Kitaguchi}, {Kolodziejczak}, {Latronico}, {Liodakis}, {Maldera}, {Manfreda}, {Marinucci}, {Marscher}, {Marshall}, {Massaro}, {Mitsuishi}, {Mizuno}, {Negro}, {Ng}, {O'Dell}, {Omodei}, {Oppedisano}, {Papitto}, {Pavlov}, {Peirson}, {Perri}, {Pesce-Rollins}, {Petrucci}, {Possenti}, {Puccetti}, {Ramsey}, {Rankin}, {Roberts}, {Romani}, {Sgr{\`o}}, {Slane}, {Spandre}, {Swartz}, {Tamagawa}, {Tavecchio}, {Taverna}, {Tawara}, {Tennant}, {Thomas}, {Trois}, {Turolla}, {Vink}, {Xie}, \& {Zane}}]{vmp2024}
{Veledina}, A., {Muleri}, F., {Poutanen}, J., {et~al.} 2024{\natexlab{a}}, Nature Astronomy, 8, 1031

\bibitem[{{Veledina} {et~al.}(2024{\natexlab{b}}){Veledina}, {Poutanen}, {Bocharova}, {Di Marco}, {Forsblom}, {La Monaca}, {Podgorn{\'y}}, {Tsygankov}, {Zdziarski}, {Ahlberg}, {Green}, {Muleri}, {Rhodes}, {Bianchi}, {Costa}, {Dov{\v{c}}iak}, {Loktev}, {McCollough}, {Soffitta}, \& {Sunyaev}}]{vpb2024}
{Veledina}, A., {Poutanen}, J., {Bocharova}, A., {et~al.} 2024{\natexlab{b}}, \aap, 688, L27

\bibitem[{{Viironen} \& {Poutanen}(2004)}]{vp2004}
{Viironen}, K. \& {Poutanen}, J. 2004, \aap, 426, 985

\bibitem[{{Wang} {et~al.}(2025){Wang}, {Marino}, {Zhang}, {Coti Zelati}, {Wang}, {Mao}, {Hou}, {Russell}, {Alabarta}, {Rea}, {Feng}, {Tao}, {Zhang}, {Pan}, \& {Yuan}}]{wmz2025}
{Wang}, B.~T., {Marino}, A., {Zhang}, G.~B., {et~al.} 2025, ATel, 17324, 1

\bibitem[{{Wijnands} \& {van der Klis}(1998)}]{wvdk1998}
{Wijnands}, R. \& {van der Klis}, M. 1998, \nat, 394, 344

\bibitem[{{Wilms} {et~al.}(2000){Wilms}, {Allen}, \& {McCray}}]{wam2000}
{Wilms}, J., {Allen}, A., \& {McCray}, R. 2000, \apj, 542, 914

\bibitem[{{Zdziarski} {et~al.}(2020){Zdziarski}, {Szanecki}, {Poutanen}, {Gierli{\'n}ski}, \& {Biernacki}}]{zsp2020}
{Zdziarski}, A.~A., {Szanecki}, M., {Poutanen}, J., {Gierli{\'n}ski}, M., \& {Biernacki}, P. 2020, \mnras, 492, 5234

\bibitem[{{Zhang} {et~al.}(2025){Zhang}, {Santangelo}, {Xu}, {Feng}, {Lu}, {Chen}, {Ge}, {Nandra}, {Wu}, {Feroci}, {Hernanz}, {Liu}, {He}, {Wang}, {Jiang}, {Cui}, {Yang}, {Wang}, {Li}, {Li}, {Du}, {Liu}, {Meng}, {Wen}, {Zhang}, {Ma}, {Li}, {Li}, {Qi}, {Sun}, {Luo}, {Liu}, {Liu}, {Zhang}, {Luo}, {Zhu}, {Zhao}, {Sun}, {Yang}, {Wu}, {Jiang}, {Shi}, {Liu}, {Xu}, {Yang}, {Zhang}, {Han}, {Gao}, {Huo}, {Zhang}, {Wang}, {Zhao}, {Wang}, {Li}, {Bao}, {Liu}, {Wang}, {Wang}, {Wang}, {Wang}, {Wang}, {Ding}, {Sheng}, {Qiang}, {Yan}, {Liu}, {Wu}, {Liu}, {Chen}, {Zhang}, {Liu}, {Altmann}, {Bechteler}, {Burwitz}, {Fiorini}, {Friedrich}, {Meidinger}, {Strecker}, {Baldini}, {Bellazzini}, {Bonino}, {Frass{\`a}}, {Latronico}, {Maldera}, {Manfreda}, {Minuti}, {Pesce-Rollins}, {Sgr{\`o}}, {Tugliani}, {Pareschi}, {Basso}, {Sironi}, {Spiga}, {Tagliaferri}, {Tykhonov}, {Paltani}, {Bozzo}, {Tenzer}, {Bayer}, {Tuo}, {Liu}, {Zhang}, {Cai}, {Liu}, {Chen}, {Wang}, {He}, {Chen}, {Qiu}, {Zhang}, {Feng}, {Zhu}, {Zhou}, {Zheng}, {Song},
  {Wang}, {Jia}, {Jiang}, {Li}, {Zhao}, {Guan}, {Zhang}, {Li}, {Huang}, {Liao}, {You}, {Zhang}, {Wang}, {Wang}, {Ou}, {Hu}, {Shi}, {Cui}, {Jiang}, {Cheng}, {Li}, {Xu}, {Zane}, {Bambi}, {Bu}, {Dall'Osso}, {Rosa}, {Gou}, {Guillot}, {Ji}, {Li}, {Mao}, {Patruno}, {Stratta}, {Taverna}, {Tsygankov}, {Uttley}, {Watts}, {Wu}, {Xu}, {Yi}, {Zhang}, {Zhang}, {Zhao}, \& {Zhou}}]{zsx2025}
{Zhang}, S.-N., {Santangelo}, A., {Xu}, Y., {et~al.} 2025, Science China Physics, Mechanics, and Astronomy, 68, 119502

\end{thebibliography}

\begin{appendix}
\nolinenumbers
\twocolumn

\section{Type-I X-ray burst observed by {\ixpe}}

\label{app:burst}

\begin{figure}[h]
   \centering
   \includegraphics[width=0.9\hsize]{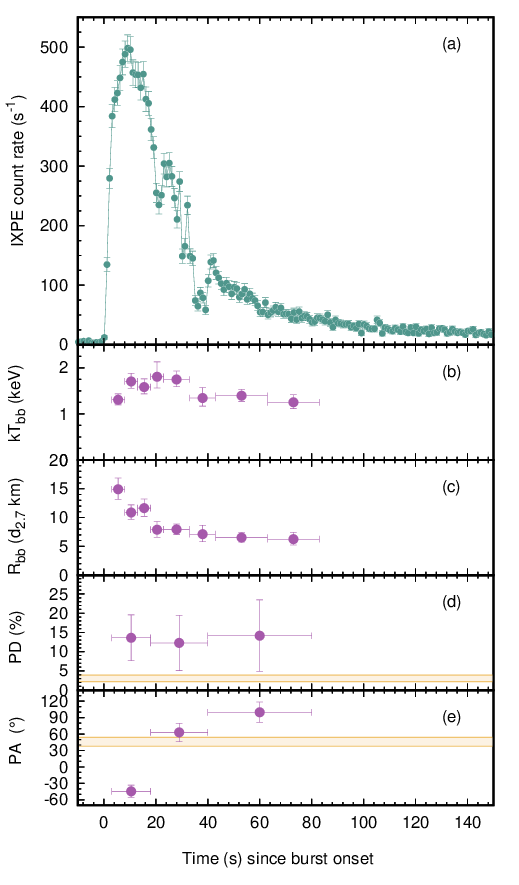}
      \caption{Light curve binned in 1~s-long intervals (panel a), burst temperature (b), effective radius (c), PD (d) and PA (e) observed by the three {\ixpe} DUs during the type-I X-ray burst that occurred on MJD~60917.62751. Yellow shaded regions mark the 90\%~c.l. intervals of PD and PA measured during the outburst decay (interval T1).}
         \label{fig:burst}
   \end{figure}

Figure~\ref{fig:burst} shows the light curve of the type-I X-ray burst observed by {\ixpe} on MJD~60917.62751 (UTC, not corrected to the Solar System barycentre). Modelling the light curve with an exponential decay indicates a time-scale of  $17.5 \pm 0.8$~s, possibly due to ignition in an almost pure He environment. The decay features two dips at $\sim 25$~s and $\sim 50$~s after the burst peak, which resemble the complex evolution seen during a bright burst observed by {\nicer} in 2019 and ascribed to the local Eddington limit of hydrogen and helium layers of the NS envelope \citep{bjg2019}. However, the {\ixpe} spectrum observed during a 15~s-long interval centred at the peak could be well fitted by an absorbed blackbody with a temperature of $kT_{\rm bb}=1.55\pm0.08$~keV and a radius of $(12.0\pm0.7)\,d_{2.7}$~km, which implies a bolometric flux of $(1.2\pm0.3)\times10^{-7}$~\fluxcgs, i.~e. roughly half that reported by \citet{bjg2019} for the photospheric radius expansion burst observed in 2019.

\begin{figure}[ht!]
   \centering
   \includegraphics[width=\hsize]{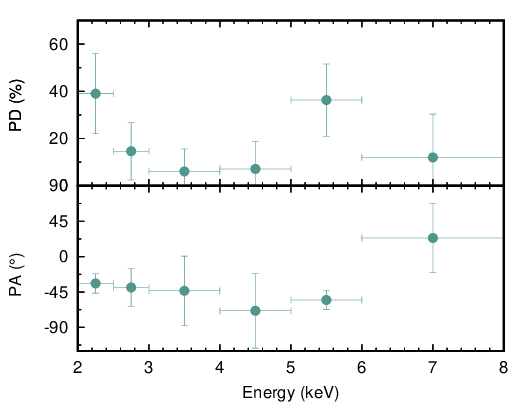}
      \caption{Polarisation degree (top panel) and angle (bottom panel) observed by {\ixpe} during a 15~s-long interval centered at the burst peak. }
         \label{fig:burst_endep}
   \end{figure}

Panels (b) and (c) in Fig.~\ref{fig:burst} show the evolution of the blackbody temperature and radius, respectively, in intervals of length ranging between 5 and 20~s. The effective radius decreases steadily during the burst decay, but the statistics do not allow us to study the evolution with the time resolution required to assess the possibility of a photospheric radius expansion \citep[see, e.g., Fig.~3 in][]{bjg2019}. Panel (d) and (e) in the same figure show the PD and PA observed during the burst decay, whereas Fig.~\ref{fig:burst_endep} shows the energy dependence of the burst properties observed during the same 15~s-long interval centred at the burst peak considered earlier to measure the peak burst energetics.

\section{Spin and orbital evolution}
\label{sec:longtermevol}

\begin{figure}[h!]
   \centering
   \includegraphics[width=\hsize]{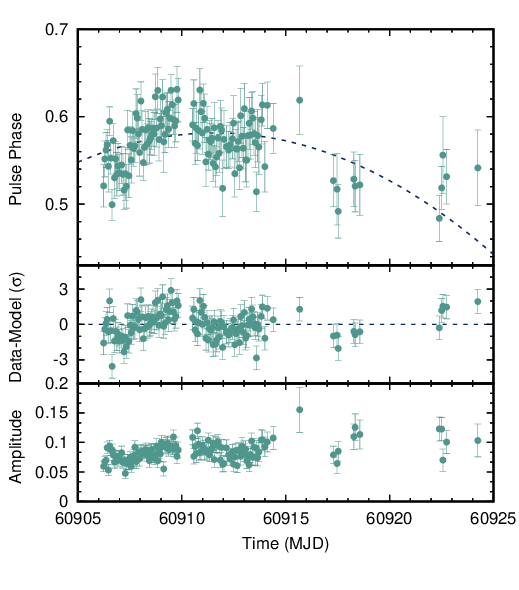}
      \caption{Pulse phase $\phi_1$ measured in 5~ks-long intervals of {\ixpe} data preliminarily corrected for the circular-orbit binary motion (top panel) and best-fitting model (dashed line; see parameters in Table~\ref{table}). Phase residuals with respect to the best-fitting model in units of the statistical uncertainties on the data points (middle panel). Fractional amplitude of the first harmonic of the pulse, $A_1$ (bottom panel).
      }
         \label{fig:timing}
   \end{figure}

We first performed an epoch-folding search on the {\nustar} time series (Obs.~Id.~81101311002 and 81101346002), after applying barycentric corrections using the position reported by \citet{bca2020} and the JPL~DE405 ephemeris. To correct for the pulsar motion in the binary system, we started from the solution found by \citetalias{ips2023} to propagate the binary and spin parameters at the epoch of the 2025 observations. We then used the preliminary timing parameters that maximised the strength of the signal in \nustar data to correct the times of arrival of X-ray photons recorded by \ixpe. We epoch-folded 5~ks-long intervals of \ixpe data and modelled the resulting pulse profiles with two sinusoidal harmonic components, 
\begin{equation}
\label{eq:pulse}
F(\phi)=\bar{F}\{1+A_1 \sin[2\pi(\phi-\phi_1)]+A_2\sin[4\pi(\phi-\phi_2)]\},
\end{equation}
where $\bar{F}$ is the average count rate, $A_i$ is the fractional amplitude and $\phi_i$ the pulse phase of the $i$-th harmonic component ($i=1,2$). After a few iterations to refine the orbital parameters, we obtained the pulse phase evolution shown in the top panel of Fig.~\ref{fig:timing}. There, the bottom panels show the residuals with respect to the best-fitting model with parameters listed in Table~\ref{table}, and the fractional amplitude of the first harmonic of the decomposition used to model the pulse profiles. Including a quadratic component $\ddot{\phi}$ significantly improved the description of the phase evolution ($\chi^2=165.5$ for 126 d.o.f.), yet timing noise suggests caution in interpreting it as accretion-driven spin-up \citep{hpc2008}.

\begin{figure}[t!]
   \centering
   \includegraphics[width=0.95\hsize]{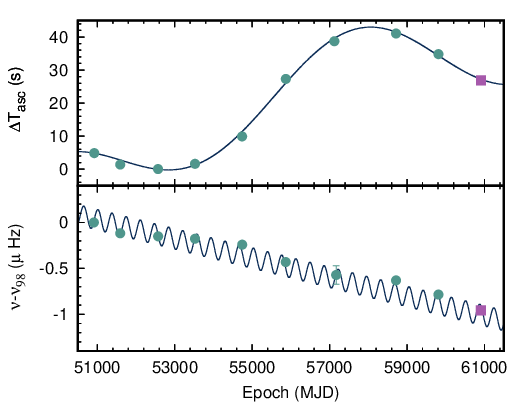}
      \caption{ Long-term evolution of the lag of the epoch of passage of the ascending node with respect to those predicted by a constant orbital period solution  (top panel) and of the spin frequency (bottom panel) of {\saxj}. Purple squares mark {\ixpe} measurements, whereas green dots mark previous measurements (\citetalias{ips2023}). Solid lines show the best-fitting solutions (see text for details).
      }
         \label{fig:spinorbevol}
   \end{figure}

We used the measured spin and orbital parameters (see Table \ref{table}) to extend the assessment of the long-term evolution of {\saxj} since 1998. The value of $T_{\rm asc}$ lags by $\Delta T_{\rm asc}=26.8\pm0.2$~s the epoch predicted according to a solution with constant orbital period $P_{\rm ref}=7249.156980$~s and reference epoch of ascending node passage $T_{\rm ref}=52499.9602472$~MJD (see \citetalias{ips2023} and references therein). Similar to \citetalias{ips2023}, we modelled the long-term evolution of the $\Delta T_{\rm asc}$ values (Fig.~\ref{fig:spinorbevol}; top panel) with an oscillation with period $P=8150\pm720$~d ($\sim 22$~yr) and amplitude $A=12.4\pm1.3$~s superimposed on an orbital period shrinkage at a rate $(-4.7\pm1.8)\times10^{-13}$~s~s$^{-1}$ (see Eq.~3 of \citetalias{ips2023}). Large residuals, yet evenly distributed around the best-fitting model, lead to a statistically unacceptable fit $\chi^2=152$ (for four d.o.f.). Other authors obtained similar results (e.g., \citetalias{ips2023}). To give reasonable estimates of the uncertainties on the fitted timing parameters, the uncertainties quoted above and in Table~\ref{table} are obtained after magnification of the error on each fitted point by a common factor such that the resulting fit reduced $\chi^2$ is equal to unity. Nevertheless, we strongly suggest caution in interpreting the uncertainties in the parameters quoted in Table~\ref{table}. We also note that the available dataset covers just one cycle of the putative $\sim22$~yr oscillation in the epoch of ascending node passage, clearly requiring a continuation of the monitoring to assess the binary long-term evolution.

The measured spin and orbital parameters confirm the long-term evolution of {\saxj}. Despite the presence of large residuals, modelling the evolution of the measured epoch of passage of the pulsar at the ascending node (purple square in the top panel of Fig.~\ref{fig:spinorbevol}) suggests a ${\approx} 12$~s amplitude oscillation with a ${\approx} 22$~yr period, similar to previous observations \citep[][I23 hereafter]{ips2023}. Similarly, the spin frequency keeps a long-term spin-down at an average rate of $\dot{\nu}=(-1.21\pm0.09)\times10^{-15}$~Hz~s$^{-1}$, superimposed on a yearly oscillation due to the difference between the best-fitting position and that determined by \cite{hpc2008} (see bottom panel of Fig.~\ref{fig:spinorbevol}). 
 
For consistency with spin frequency measurements used in the past to evaluate the spin long-term evolution, we measured the 2025 average frequency of {\saxj} by correcting {\ixpe} data with the source position by \citet{hpc2008}, and neglecting any quadratic component $\ddot{\phi}$ in the model. Modelling this and previous frequency measurements (see \citetalias{ips2023} and references therein) with a constant spin-down model superimposed with a yearly sinusoidal oscillation due to corrections to the source position (see Eq.~2 in \citetalias{ips2023}) gave ($\chi^2=21.5$ for 6 d.o.f.) $\dot{\nu}=-(1.21\pm0.09)\times10^{-15}$~Hz~s$^{-1}$ and R.A.(J2000) $=18^{\rm h}08^{\rm m}27\fs66\pm0\fs02)$, dec (J2000) =$-36\degr58\arcmin44\farcs4\pm0\farcs6)$, compatible with previous estimates (see \citealt{bca2020}, \citetalias{ips2023}). Deviations of the actual frequency from the average value measured during each of the outbursts, as well as pulse phase timing noise \citep[see, e.g.,][]{hpc2008}, could explain the large residuals. Similar to the case of the evolution of the ascending node epoch, we quote uncertainties on the parameters describing the spin evolution as those obtained by magnifying uncertainties on single data points by the factor required to obtain a reduced $\chi^2=1$.

\section{Modelling of Stokes parameters phase-resolved variability}
\label{sec:pulsemodel}

To give a preliminary modelling of the pulse profiles of the Stokes parameters $(I,Q,U)$ of {\saxj} (see Fig.~\ref{fig:pulse}), we used the same approach as \citetalias{pdmp2025}. We described the dependence of the Stokes parameters on the pulse phase using the analytical relations of \citet{pb2006} and \citet{p2020}. We assumed two point-like spots at the surface of a spherical NS spinning at 401~Hz, with a mass of $1.4\,M_{\odot}$ and an equatorial radius of 12~km, emitting radiation with a spectrum described by a power-law index $\Gamma=1.8$, an anisotropy parameter $h$ and energy-independent polarisation $p$ (see Fig.~9 of \citetalias{pdmp2025} for a sketch of the geometry). The observed values of $(I,Q,U)$ are normally distributed \citep{sft2023} and we obtained a best-fit ($\chi^2=45.1$ for 36 d.o.f., see solid lines in Fig.~\ref{fig:pulse}) for a system inclination $i=60{\degr}\pm1{\degr}$, co-latitudes of the two spots $\theta_1=6^{\,{+2}}_{\,-1}$~deg and $\theta_2=177{\degr}\pm 1{\degr}$, longitude of the first spot $\phi_1=13^{\,+11}_{\,-16}$~deg, longitude separation of the two spots $\Delta\phi=\phi_2-\phi_1=108^{\,{+16}}_{\,-12}$~deg, position angle of the pulsar angular momentum $\chi_{\rm p}=41{\degr}\pm5{\degr}$, PD of the emission of each spot of $p=3.2\%\pm0.6\%$, and emission anisotropy parameter $h=-0.4^{\,{+0.2}}_{\,{-0.1}}$. In this geometry, photons from the primary spot are seen at all phases, whereas the secondary is occulted in the phase range $0$--$0.3$. Solid curves in Fig.~\ref{fig:pulse} show the phase-resolved variability of the Stokes parameters, the PD and the PA according to this model.

\begin{figure}[t!]
\includegraphics[width=0.96\hsize]{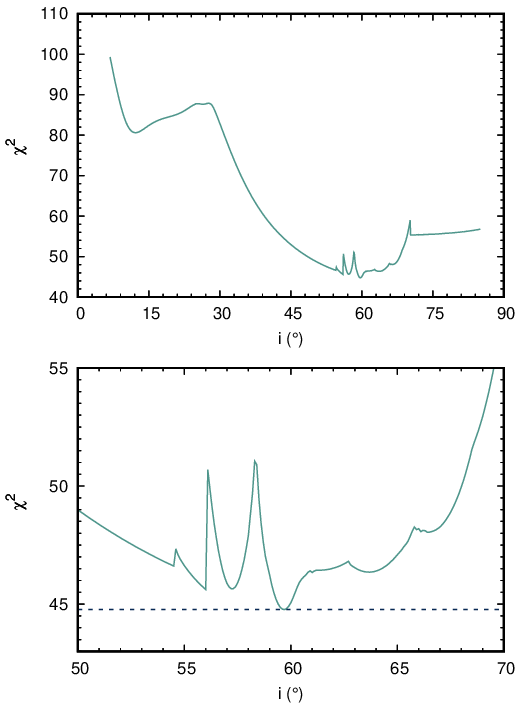}
\caption{Best-fit $\chi^2$ for the model described in Appendix \ref{sec:pulsemodel} fitted to the observed phase-resolved Stokes parameters (see Fig.~\ref{fig:pulse})  as  a function of inclination. The top panel shows the values in a broader range of inclinations, while the bottom panel zooms in on an interval around the global minimum. }
\label{fig:inclinationscan}
\end{figure}

Owing to the data quality and the multiple parameters, the $\chi^2$ distribution in the parameter space supports numerous local minima of comparable value. Figure~\ref{fig:inclinationscan} shows best-fit $\chi^2$ values obtained through least-squares fitting, by fixing the angle of inclination on a grid of points separated by $\delta i = 0.1{\degr}$, and leaving the other parameters free. Several local minima are found between $i\simeq 54^{\degr}$ and $67^{\degr}$. The $\chi^2$ at those minima ranges between 46 and 50, only slightly higher than our best fit. This suggests that current data constraints on the inclination have to be taken with caution. All models require a small value of $\theta_1$ and an almost antipodal $\theta_2$, with angular distances between the two of $150\degr$--$180{\degr}$, though. Despite the simplistic assumptions, this indicates that the nearly-absent phase variability of the Stokes parameters can be ascribed to a low co-latitude for the main, more visible hotspot. 

\begin{figure*}[h!]
   \centering
   \includegraphics[width=\textwidth]{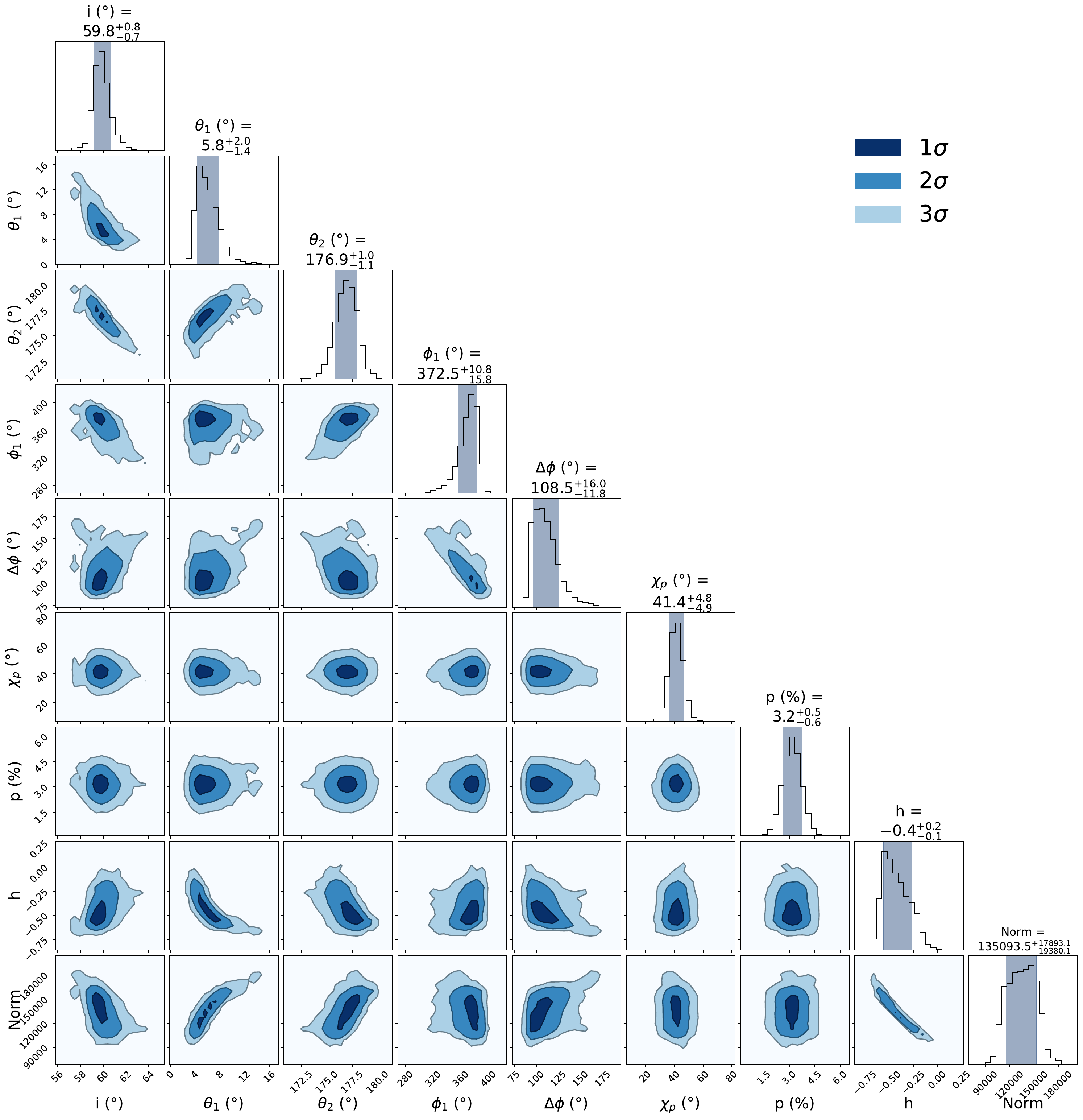}
      \caption{Corner plot showing the most likely distributions for the parameters of the pulse-profile model. 68, 95, and 99.7\% confidence levels are indicated as coloured lines encompassing differently shaded areas. The degeneracies observed between different parameters, such as between the inclination and the emission co-latitudes of the two hotspots, or between $h$ and the total normalisation, are what is expected from our idealised model. Each parameter's distribution graph also shows the 68\%-probability region around its median as a shaded rectangle. 
      }
         \label{fig:cornerplot}
   \end{figure*}

To avoid local minima, we employed a Monte-Carlo Markov Chain (MCMC) maximum-likelihood exploration over the whole parameter space,\footnote{As implemented by the LMFIT package \citep{lmfit}.} starting from uniform priors. The best-fit values quoted above and their uncertainties are the median and 68\% credible intervals, respectively. Figure~\ref{fig:cornerplot} shows the respective corner plot after running the MCMC with 200 walkers.

\end{appendix}
\end{document}